# A Privacy Scheme for Monitoring Devices in the Internet of Things

Zygmunt J. Haas[1] and Ashkan Yousefpour[2]

[1] Wireless Networks Laboratory, Cornell University, Ithaca NY 14853, USA
`zhaas@cornell.edu`
[2] Department of Computer Science, University of Texas at Dallas, Richardson TX 75080, USA
`ashkan@utdallas.edu`

**Abstract.** Sufficiently strong security and privacy mechanisms are prerequisite to amass the promising benefits of the IoT technology and to incorporate this technology into our daily lives. This paper introduces a novel approach to privacy in networks, an approach which is especially well matched with the IoT characteristics. Our general approach is based on continually changing the identifying attributes of IoT nodes. In particular, the scheme proposed in this work is based on changing the IoT nodes' IP addresses, and because the changing patterns of the IP addresses appear random to a non-intended observer, an adversary is unable to identify the source or destination of a particular transmission. Thus, packets that carry information generated by a particular node cannot be linked together. The scheme offers additional security benefits, including DoS mitigation, is relatively easy to implement, and requires no changes to the existing networking infrastructure. We discuss the details of the implementation of the scheme and evaluate its performance.

**Keywords:** Privacy, Anonymity, IoT, Security, IP Address Hopping

## 1 Introduction and Motivation

To amass the promising benefits of the Internet of Things (IoT) technology, a number of technical challenges have to be overcome, with security being a major such a challenge. Without sufficient degree of security and privacy of information, users will not adopt this new trend that promises to intimately integrate into their lives. It is generally believed that security of Internet of Things is a significantly more challenging problem than the security of today's Internet. First, the number of devices in the IoT increases exponentially and many of these devices will operate unattended, thus more time might pass without a successful attack being detected. Moreover, all the malware that already exists today in the Internet, become viable threats to the small-print IoT devices, incapable of running complex security protection software. Furthermore, a successful attack on IoT devices, such as medical devices, baby-monitoring equipment, smart stove, and house alarm systems, creates potential for severe and immediate danger to their users (e.g., resulting in injury or death), a different type of danger than we are used to with typical Internet malware, such as theft of information.

There have been a number of solutions proposed in the literature that preserve privacy for IoT networks (e.g., [1] – [3]). However, as Internet transmissions require explicit disclosure of source/destination IP addresses, these schemes cannot hide the identity of the IoT nodes, thus allowing the adversaries to learn about the IoT nodes simply by observing the IP addresses in the packets' headers. In contrast, our proposed scheme, aims to actively obfuscate the IP address of a node by allowing the IP address of the node to change frequently (i.e., "IP address hopping"), thus creating uncertainty for adversaries of who is the source/destination of a transmission, while still allowing the packets to be correctly routed to the destination within the Internet.

As an example, consider a hospital facility in which numerous patients are hooked up to medical sensing IoT devices (e.g., EKG, SPO2, GSR, BP, temperature, etc), together creating an IoT network. The sensors' readings are continuously acquired, packetized, and transmitted to the medical information collection station for processing, archival, and possibly alerting medical personnel of emergency care needed. Such transmissions, being IP-routed, contain the IP addresses of the source device – the IoT sensor of the patient. Typically, such information would also include the identity of the patient. As all the packets originating from the same IoT device would carry the same IP address, an adversary can assemble the medical record of a patient by collecting subsequent packets. In other words, the IP addresses create an index that links all the transmissions together.

Another example could be collection of electricity reading from electric meters. The importance of privacy of such information is well acknowledged, as it could be used by thieves to determine that the house occupants are away and, thus, the house may be subject to a burglary. Of course, a series of readings put together would tell whether the electricity reading decreased in a particular time period, indicative of the occupants being away. Our scheme can preserve the privacy of such information by severing the link between the electricity readings, as well as the readings' link to any ID of a residence.

Using the proposed-here scheme, the IP addresses of subsequent transmissions of each IoT device would be changed in some unpredictable (yet deterministic) pattern, so that the adversary would not be able to use the IP addresses as a linking index of the transmissions. In other words, the adversary will see a massive collection of readings, but will not be able to attribute any reading to a single source (e.g., patient or house, in the previous examples). Of course, the receiver would need to generate a corresponding sequence of the IP addresses, so that the receiver can properly collect together the received information. We further note that, as the pattern of IP addresses is unique to a particular device, there is no need to include the encrypted patient's ID in the packets, as the IP address pattern already identifies a particular IoT device to the receiver (but not to the adversary). In other words, the IP address pattern serves as an ID of the IoT device. Furthermore, an attempt to associate a patient with an IP address of his IoT devices would also be fruitless.

## 2   The Basics of the Scheme

The proposed scheme is useful for information privacy protection in a scenario where a large number of IoT devices transmit similar monitoring (e.g., telemetry) data. More particularly, each transmitted data packet, standing by itself and without association with a particular user, would be useless to an attacker, while either (1) collection of large amount of data coming from a particular user, or (2) association of the data with a particular user, would constitute breach of information privacy. The example of a hospital with large number of the same type of medical sensors would correspond to such a scenario. Similarly, the example of electric meter information from numerous houses in a neighborhood would also present such a scenario.

The basic setup of our scheme includes three nodes, the *IoT node* whose information privacy we intent to protect, the device that communicates with the IoT node, which we refer to as the *corresponding node (CN)*, and a trusted node that controls the operation of the scheme, which we refer to as the *central node*. In a general scenario multiple IoT nodes communicate with multiple corresponding nodes.

The IP address hopping is achieved by a *pseudorandom number generator* that is embedded in a function referred to here as the *Tracking Function (TF)*. The parameters of the *TF* are shared by the IoT node and the authorized CNs. (Note that the *TF* itself does not need to be secret) The *TF* continually generates, what appear to an arbitrary observer, random addresses. We emphasize that although the output of the *TF* seems random, the operation of the function is deterministic; i.e., anyone who observes the output of the *TF*, even for a long time, cannot predict its future values; but whoever holds the *parameters* (including the input) of the *TF* can replicate the output deterministically.

An IoT node uses the random addresses as its actual addresses as they are generated by the *TF*. When an authorized CN desires to communicate with the IoT node, (authorized CN is in possession of the *TF* parameters), it uses the valid (i.e., the current) address generated by the *TF* as the destination address of its transmission. Similarly, transmission from the IoT node uses as the source address the currently generated output based on the *TF*. The IoT node and the CNs generate the IoT node's current IP address every $\zeta$ seconds. Of course, for the scheme to operate properly, some degree of synchronization of the *TF* at the IoT node and the CNs is required – we discuss this in more details later.

The role of the central node is mainly to perform the coordination functions: authenticate the CNs, distribute the TF parameters, and aid in clock synchronization. The central node, the IoT nodes, and the CNs do not have to reside on the same network or even be close to each other. We assume here that the IoT node is static and does not migrate to a new subnet while the scheme is operating, although the scheme could be easily extended to support mobile operation as well.

Our scheme does not introduce additional header information for its operation and it can be incrementally deployed in networks; furthermore, the scheme is compatible with IPv6 addressing. There is no change required for the operation of routing and switching. The required changes to the IP protocol are mostly in the end nodes (the

IoT and the CN nodes). If the changes in IP address are sufficiently fast, the scheme could also be used for DoS mitigation at the IoT node.

An alternative scheme would be to implement end-to-end encryption on each of the IoT devices' information flows. Although this would protect the information privacy, we suggest here that the IP address hopping provides significant advantages over encryption. In what follows, we explain why.

If end-to-end encryption were to be implemented, it is clear that multiple keys (probably one key per an IoT device) would need to be maintained. Therefore, some node ID would have to be transmitted in the clear to allow the receiver to choose the proper decryption key. (In fact, the IP address could be such a node ID used to choose the proper key.) As such, the attacker would be able to associate packets with a particular ID, risking loss of privacy. On the other hand, in the proposed scheme, no node ID needs to be transmitted; indeed, even the IP address of the node cannot be interpreted as a node ID, as it is continually changes (even if an attacker is able to associate an IP address with a particular device, such an association would be very short-time living with very limiting privacy consequences). Thus, we maintain that, for the assumed communication scenario, our scheme provides advantageous information privacy scheme, compared with plain encryption.

Furthermore, the proposed scheme avoids the need to maintain the encryption keys and the necessity to periodically rekey the nodes. Finally, the overhead associated with encryption/decryption is eliminated too, which is of particular benefit for resource-constrained devices.

## 2.1 Threat Model

We assume that an adversary can mount passive attacks, such as network scanning and eavesdropping to collect information carried by the packets (including the header information), to assemble information from packets, so as to obtain protected information sent by the IoT nodes (i.e., violating privacy). An attacker can eavesdrop on all connections. In particular, a passive attacker can obtain the current IP address of the IoT node and launch attacks on the IoT node (i.e., becoming an active attacker). We assume that network infrastructure is reliable and not malicious; but may impose delay and packet loss. We further assume that CNs are not malicious and that the central node is a trusted node.

## 2.2 The Tracking Function

In order to generate the IP addresses at the IoT node, we use the timestamp (a sequence that is linearly increasing) as the input to a pseudo-random number generator (PRNG). The timestamp of the IoT node is one of the parameters that is kept secret in our scheme and is in the possession of the secret-sharing nodes. The PRNG, on the other hand, is publicly known; however, without knowing the timestamp and the other parameters the output is unpredictable. In general, any hash function that satisfies the following characteristics, can be used as the scheme's PRNG:

- The function must be one-way secure, meaning that by watching the past values, one cannot guess the parameters of the function.
- The function must be unpredictable; meaning that by watching past values, one cannot predict any future values of the function.
- The function outputs should be randomly distributed on any time scale (at least on a sufficiently long time scale).

The IP address of IoT node is generated by feeding the timestamp to the *Tracking Function*, which is based on PRNG as follows:

$$IP = TF(timestamp) = BA + H_x(timestamp) , \qquad (1)$$

where *TF* denotes the *Tracking Function*, *BA* represents the base address of the IoT node's subnet (e.g. '129.110.242.0' without '/24'), and $H_x$ denotes using $x$ least significant bit of the output of the PRNG. $x$ is the minimum number of bits that is required for representing all the available addresses in the IoT node's subnet (*BA* and $x$ can be calculated from the IoT node's subnet address).

 We propose to use a chaotic function as the PRNG. In general, chaotic functions are highly sensitive to initial conditions and control parameters, and they appear to behave randomly, alas they are completely deterministic once the set of control parameters is known. A slight change in the input will result in a big change in the output. This property fits well with the goals of the PRNG. More specifically, we use the Hash Function Based on Chaotic Tent Maps as the PRNG of the scheme ([4]), since it has the aforementioned characteristic. By using the hash function based on a chaotic function, a third-party can neither predict the future values by watching the function, nor generate the function without having the control parameters.

 The following is a simple example that demonstrates the operation of the *Tracking Function*. We further assume that we are using IPv4 addressing scheme and that the network address of IoT is 129.110.242.0/24. We need at least 8 bits to represent the host ID portion of the IP address ($x = 8$). Table 1 shows the corresponding generated IP addresses.

**Table 1.** Output of the *Tracking Function* for 6 samples of timestamp

| Time-stamp | 8 least significant bits of PRNG output | | *Tracking Function* output |
|---|---|---|---|
| | Binary | Decimal | |
| 3000000 | 10000111 | 135 | 129.110.242.135 |
| 3000001 | 00010100 | 20  | 129.110.242.20  |
| 3000002 | 11101100 | 236 | 129.110.242.236 |
| 3000003 | 11111100 | 252 | 129.110.242.252 |
| 3000004 | 00101010 | 42  | 129.110.242.42  |
| 3000005 | 00010010 | 18  | 129.110.242.18  |

 Basically, the hash function based on the chaotic tent maps takes in an arbitrary length input $M$ and produces a $2l$-bit hash output, where $l$ is the blocks' size into which the message $M$ is broken. $n$ is the number of rounds in the function. If $M < l$, the block is padded so that the size of the message is a multiple of $l$. In our scheme, the hash function takes in the timestamp as the input M and a pair of initial binary fractions $(s_0, t_0)$, producing a hash output that is a $2l$-bit binary number. Yet we only use the required number of bits $(x)$ that is needed to represent all the available IP addresses in a subnet. The initial parameters $(s_0, t_0)$ could be chosen in different ways, but for a good perturbation we use here $(s_0, t_0) = (0.1010…10, 0.0101…01)$. In [4], the author showed that the hash function is resistant to target attack, free-start target attack, collision attack, semi-free-start collision attack, and free-start collision attack, as the computational complexity of these attacks are $2^l$, $2^l$, $2^{l/2}$, $2^{l/2}$, $2^{l/2}$ respectively.

After successful authentication with the central node, authorized CNs get the parameters of the *Tracking Function* from the central node. The parameters are: timestamp, $\zeta$, $l$, and subnet address of the IoT node.

## 2.3    Clock Synchronization

As discussed below, some degree of clock synchronization is required in the scheme to guarantee that timestamps of the central node, the CNs, and the IoT nodes are synchronized. Clock synchronization algorithms sync two or more clocks that have a non-zero drift rate. Typically, drift rate is a very small number; but due to the high frequency of clocks, this can lead to a large difference in clocks even after a short while. The timestamp that we use in our solution, however, is different from the local clock of the operating system. The timestamp that we use is a number that increases by one every $\zeta$ seconds. The central node, after authenticating the CN, performs coarse clock synchronization with the CN, before sending the *Tracking Function* control parameters to the CN. Note that all the nodes (central node, IoT node, and CN), perform clock synchronization periodically.

Let us assume that $\eta$ is the number of times that an IP address changes in each clock synchronization period $\tau$; i.e., $\tau = \zeta \times \eta$, where $\eta$ is a parameter that reflects the accuracy of the clocks in use and is calculated based on the maximum drift rate as follows. Assume that the maximum drift rate in the system is defined by $\delta$ [sec/sec]. Usually $\delta$ is a small number (e.g., $10^{-6}$). The maximum skew between the clocks in the system after 1 second would be $2 \times \delta$ [sec]. We know that timestamp increases by one every $\zeta$ seconds. The maximum skew between two timestamps should be always kept less than one within the interval of clock synchronization (every $\tau$ seconds). This way the timestamps will always be equal, since they are integer numbers. Let $S$ denote the skew between the timestamps within $\tau$ seconds; thus we require that $S < 1$:

$$S = 2 \times \delta \times 1/\zeta \times \tau = 2 \times \delta \times \eta \longrightarrow \eta < \frac{1}{(2 \times \delta)} \tag{2}$$

There are many clock synchronization solutions in the literature that can be used for our scheme (e.g., [5],[6]). For instance, Network Time Protocol (NTP) is a low-cost solution whose accuracy ranges from hundreds of microseconds to several milliseconds ([7]). The reference [8] presents a precise relative clock synchronization protocol for distributed applications. It achieves clock precision on the order of 10 microseconds in small-scale LANs and sub-millisecond over LANs. For our experiment (Section 4), we implemented an NTP-like clock synchronization program, where the central node is an NTP server and the other nodes in the system synchronize their clock with it.

## 3      Performance Issues

### 3.1    Address Collision

If many IoT nodes in a subnet use the IP hopping scheme, there is a probability that, at some point in time, two (or more) nodes will be assigned the same IP address. This, of course, is an undesirable situation that should be avoided. In this section, we estimate the probability of such an *address collision*.

Suppose that, in a particular subnet, there are $k + h$ nodes, $k$ of which are IoT nodes and $h$ are other non-IoT nodes (e.g., assigned permanent IP addresses). Further, assume that $m$ is the total number of available IP addresses in the subnet. Then, the probability that two or more IoT nodes will be randomly assigned the same IP address (i.e., the probability of address collision) is:

$$p = 1 - \frac{(m-h) \times (m-h-1) \times \ldots \times (m-h-k+1)}{m^k} \quad (3)$$

We assumed that each IP address can be assigned by the *Tracking Function* with equal probability of $1/m$, because the *Tracking Function* is technically based on a pseudo-random number generator, thus the probability of all possible outputs is equal ([4]). The author in [4] maintains that for any $0 \leq \alpha < 1$, the distribution of $x_1 = G_\alpha(x_0)$, which is the core of the Hash Function based on Chaotic Tent Maps, for randomly chosen $0 < x_0 < 1$ is the standard uniform distribution, $\mathcal{U}(0,1)$.

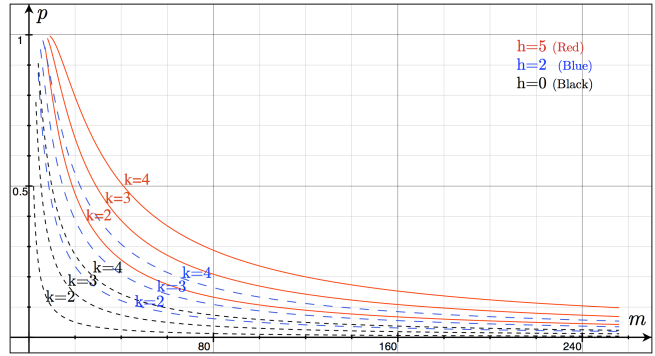

**Fig. 1.** Address collision probability as a function of address range, $m$, with $k$ IoT nodes.

Fig. 1 shows the address collision probability as a function of the address space size, $m$, for different values of $k$ and $h$ ($h + k < m < 256$). As shown in the figure, when there are only IoT nodes in the subnet (i.e., $h = 0$), the address collision probability for network sizes of $m > 40$ is negligibly small. When there are 5 normal nodes ($h = 5$) in addition to the active IoT nodes, the address collision probability is not negligible anymore. This provides guidance to the design process of such IoT subnets.

## 3.2 Packets in Transit

As discussed before, due to clock mis-synchronization and intrinsic network delays, packets arriving after a change in IP address has occurred at the IoT node, may still carry the old IP address of the IoT node and, thus, may be discarded at the destination. A mechanism is needed that will prevent or at least minimize the loss of packets in transit during the changes of network addresses. In the approach that we propose here, the IoT node continues to maintain the old IP address (together with the new one) for a short while, so that packets arriving with the old IP address after the IP address has already changed will still be accepted. Of course, the duration of time when both IP addresses are in use should be short to achieve higher privacy in IoT node, as well as to reduce the probability of address collision.

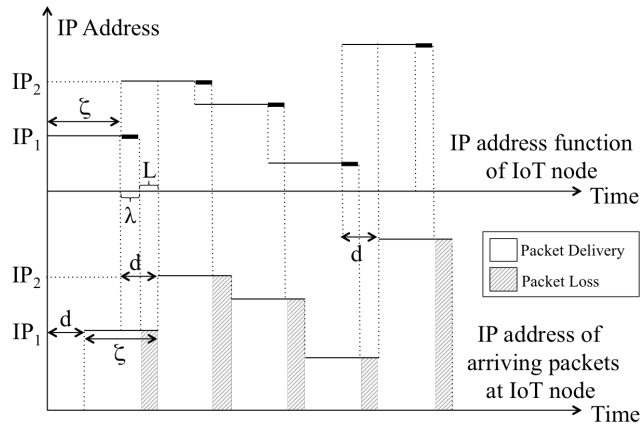

**Fig. 2.** Address Possible packet loss due to mismatch of IP addresses. Upper graph: IP address at the IoT node; lower graph: IP address of packets arriving at the IoT node.

The timing diagram explaining the scheme's operation is presented in Fig. 2. In the upper portion of the diagram presented are the assignments of the IP addresses to the IoT node as a function of time. As we can observe, initially, the IP address of $IP_1$ is assigned to the IoT node and is maintained for the period of $\zeta$, after which time the new $IP_2$ is assigned. However, $IP_1$ is kept active for an additional time $\lambda$ (the thicker line), during which time the IoT node is assigned both $IP_1$ and $IP_2$ addresses.

We now assume that the network introduces delay of $d$ to the packet sent from the CN. The lower portion of Fig. 2 displays the IP addresses of packets arriving at the IoT node. In this example, we assume that the only cause of mis-synchronization of the IP addresses is the network delay (i.e., that the clocks are perfectly synchronized). We see that the IP addresses of packets arriving at the IoT node follow exactly the IP addresses assigned at the IoT node (upper portion of Fig. 2), but they are delayed by $d$. In this example, $\lambda < d$, so some packets arrive at the IoT node after the old IP address, $IP_1$, has already been dropped (after the extra time $\lambda$); such packets are lost (the loss is marked in darker vertical spaces in the lower portion of Fig. 2). It is easy to observe that if $\lambda > d$, then no packet loss would occur. Next, we present a simple analysis of the packet loss.

Let us consider a long time interval $T$ and, for simplicity, assume that $T$ is a multiple of $\zeta$; i.e., $T = c \cdot \zeta$, for some integer $c$. We further assume that the generation rate of packets by the CN is Poisson with rate $\gamma$. Since we assume that the only source of packet loss is the mismatch in IP addresses (i.e., no network losses), the total arrival of packets to the IoT node is also Poisson with rate $\gamma$. Then the average packet loss is:

$$E[Packet\ Loss] = \frac{E[\text{number of lost packets in the interval } T]}{E[\text{total number of sent packets in the interval } T]} = \frac{E\left[\begin{array}{c}\text{total number}\\\text{of sent packets}\\\text{in the interval } T\end{array}\right] - E\left[\begin{array}{c}\text{number of}\\\text{received packets}\\\text{in the interval } T\end{array}\right]}{E[\text{total number of sent packets in the interval } T]}$$

$$= \frac{\gamma \cdot c \cdot \zeta - \gamma \cdot c \cdot (\zeta - \min(0, d-\lambda))}{\gamma \cdot c \cdot \zeta} = \frac{\min(0, d-\lambda)}{\zeta} = \frac{\min(0, L)}{\zeta}, \quad (4)$$

where we labeled $d - \lambda \equiv L$. Thus, the measure of probability of loss is $L$. To minimize the probability of loss, either $\lambda \cong d$ or $\zeta \gg d - \lambda$. The first case requires the knowledge of the value of $d$, which typically has a non-stationary distribution. Similarly, in the second case, when $d$ is large, it requires either large $\zeta$ or large $\lambda$, leading to limiting degree of achievable privacy. In either case, there is a need for a mechanism to estimate the value of $d$, which can be measured by a one-way delay measurement scheme.

### 3.3 Privacy Protection

The privacy of the scheme primarily relies on the fact that the sequence of the generated IP addresses cannot be predicted neither by anyone who does not possess the parameters of the *Tracking Function*, nor by observing the past sequence of the IP addresses. To test the temporal randomness of a function the standard method is to compute the correlation of the function at various times, i.e., the function's *autocorrelation*. We conducted experimentation with the hash function we used in our scheme, collecting the samples over sufficiently long time to calculate the autocorrelation. The experiment showed white-noise like autocorrelation (an impulse $\delta(x)$ response), demonstrating the lack of correlation in the hash function based on chaotic tent maps.

## 4 Experimentation Results

In this section, we provide some results of the experimental implementation of the scheme. We used three machines as the main components of the scheme. One machine served as the IoT node, one as the central node, and one as the CN that communicated with the IoT node. In order to evaluate the behavior of the basic scheme, we experimented with the scheme over a local-area (UTD, in Richardson, TX) network, as well as over a wide-area network, where the CN resided at Cornell University, in Ithaca, NY. The goal was to understand the performance as a function of different settings of the scheme, with drastically different distributions of the network delays. The results are summarized in Table 2 (local-area network) and Table 3 (wide-area network), for two values of λ=0.3s and λ=0.8s and $\zeta = 1, 2, 3, 4,$ and 8 seconds.

**Table 2.** Packet loss (%) for different values of $\zeta$ and $\lambda$ (Experiment over local-area network)

| | (a) $\lambda = 0.3$s | | | | | | (b) $\lambda = 0.8$s | | | | |
|---|---|---|---|---|---|---|---|---|---|---|---|
| $\zeta$(sec) | 1 | 2 | 3 | 4 | 8 | $\zeta$(sec) | 1 | 2 | 3 | 4 | 8 |
| Mean | 2.22 | 0.87 | 0.66 | 0.39 | 0.29 | Mean | 1.20 | 0.86 | 0.95 | 0.80 | 0.36 |
| 95% CI | [1.93,2.51] | [0.83,0.9] | [0.58,0.73] | [0.35,0.42] | [0.28,0.3] | 95% CI | [1,1.39] | [0.8,0.93] | [0.8,1.07] | [0.66,0.92] | [0.3,0.41] |
| Min | 0.18 | 0.67 | 0.30 | 0.17 | 0.24 | Min | 0.18 | 0.29 | 0.54 | 0.38 | 0.21 |
| Max | 4.09 | 1.16 | 1.31 | 0.58 | 0.37 | Max | 3.08 | 1.92 | 2.11 | 2.07 | 0.87 |

**Table 3.** Packet loss (%) for different values of $\zeta$ and $\lambda$ (Experiment over wide-area network)

| | (a) $\lambda = 0.3$s | | | | | | (b) $\lambda = 0.8$s | | | | |
|---|---|---|---|---|---|---|---|---|---|---|---|
| $\zeta$(sec) | 1 | 2 | 3 | 4 | 8 | $\zeta$(sec) | 1 | 2 | 3 | 4 | 8 |
| Mean | 5.60 | 1.56 | 1.37 | 0.96 | 0.88 | Mean | 1.83 | 0.98 | 1.30 | 0.83 | 0.52 |
| 95% CI | [4.68,6.51] | [1.43,1.67] | [0.45,2.29] | [0.04,1.87] | [0,1.79] | 95% CI | [1.67,2] | [0.06,1.9] | [0.38,2.2] | [0,1.74] | [0,1.43] |
| Min | 0.80 | 0.88 | 0.27 | 0.38 | 0.36 | Min | 0.61 | 0.21 | 0.40 | 0.37 | 0.18 |
| Max | 9.98 | 2.21 | 3.18 | 2.14 | 1.46 | Max | 2.56 | 1.57 | 2.45 | 1.21 | 1.02 |

In our implementation, the IoT node resides in a network of size 256. We used the following parameters for the Tracking Function: $l = 16$, $n = 75$, $x = 8$ and $(s_0, t_0) = (0.10101010, 0.01010101)$.

As we can see in Table 2, in local-area networks, for $\zeta > 2$s, the packet loss is smaller than 1%. To achieve similar packet loss, in the wide-area network, it is required that $\zeta > 4$s (Table 3). In the experiment in local-area network, most of the losses occur due to delays of running the code, and in particular, due to the delay required for changing the IP addresses in a Linux machine.

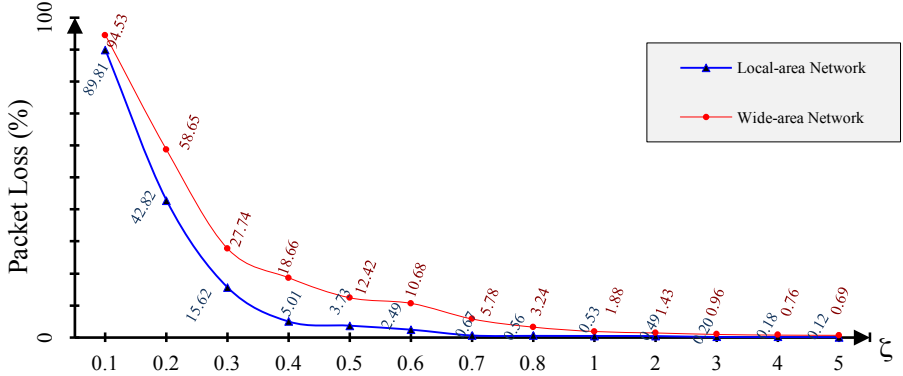

**Fig. 3.** Average packet loss (%) for different $\zeta$ (sec)

Fig. 3 shows the average packet loss for a range of values of the parameter $\zeta$, where $\lambda = 0.2 \cdot \zeta$. This experiment was done both over the LAN at UTD and also over a WAN (where the CN was at Cornell University). The same parameters for the *Tracking Function* were used as before. The figure demonstrates that there is a "threshold" value of $\zeta$, below which the packet loss rapidly increase, while above the threshold the packet loss remains relatively negligible. Thus, as long as $\zeta$ is above the threshold, the packet loss is not much sensitive to the actual value of $\zeta$. In Fig. 3, this threshold is $\zeta = 0.7$s for the experiment over the local-area network and $\zeta = 1$s for experiment over the wide-area network.

## 5 Related Work

There are some related works that use address hopping technique, although either in a different manner or for a different reason. Some of these works are primarily related to the well-known concept of frequency hopping in wireless networks. The works of Shi et al. ([9], [10]) discusses port and address hopping for active cyber defense (generally DoS). In their work, privacy is not always preserved, as hopping is not done continuously. In their more recent work ([11]), they presented a scheme that requires Hopping Agent that is responsible for the hopping operation while the security-critical server is hidden behind it. Our approach does not require agent and is more suitable for IoT scenarios where devices are connected in different settings.

Another interesting work based on address hopping is Mirage ([12]), which is mainly designed for defending against DDoS for web applications. However, the scheme cannot be used for privacy preserving in the IoT, since, the scheme is only activated when under active attack (and only then, it hops every 5 minutes). Furthermore, it does not match the resource-constrained devices in IoT, as it requires solving puzzles, filtering by routers, and sending large size ACL files (few hundred thousand entries for small attacks) to routers, on each address change.

Similarly, a more recent related work based on address hopping by Krylov et al. ([13]) addressed DDoS attack mitigation. Their system is not scalable to the IoT networks, since several routers are required (and should support the scheme) to protect a single node. Comparably, the work [14] discusses the general idea of network address hopping, but it is not suited for IoT networks, as much information needs to be sent between two peers each time a communication needs to be established between two nodes. Also their scheme is not scalable to IoT, as the hopping is on per-packet basis (only one-to-one communications is supported). In contrast, in our scheme, only the scheme's parameters (i.e., a few numbers) are transmitted when a new CN joins, and it supports one-to-many communication, namly suitable for IoT.

## 6 Conclusion

We introduced and discussed a scheme for data privacy in IoT based on IP address hopping. The scheme is in particular useful for information privacy protection in a scenario where a large number of IoT devices transmit similar monitoring (e.g., telemetry) data. To implement the scheme, we used a hash function based on chaotic tent maps as the scheme's PRNG. We discussed and evaluated some performance aspects of the scheme, such as the IP address collisions and the degree of privacy protection. In its basic configuration, the scheme requires no changes to the existing networking infrastructure. Finally, we provided the results of our experiments with the scheme and we showed that there is a fundamental trade-off between achievable degree of privacy and the average packet loss. As noted, the scheme could be also used for location-privacy and for protection against DoS attacks. We intend to evaluate these directions in our future work.

# 7 Acknowledgement

The work of Z. J. Haas has been supported by the NSF grant numbers: CNS-1040689, ECCS-1308208 and CNS-1352880.

# 8 References

1. Beresford, A.R., Stajano, F.: Mix zones: User privacy in location-aware services. In: PERCOMW '04 Proceedings of the Second IEEE Annual Conference on Pervasive Computing and Communications Workshops. pp. 127–131 (2004).
2. Fan, Y., Lin, B., Jiang, Y., Shen, S.: An Efficient Privacy-Preserving Scheme for Wireless Link Layer Security. In: IEEE Globecom Wireless Comm. Symp. pp. 4652–4656 (2008).
3. Banerjee, D., Dong, B., Taghizadeh, M., Biswas, S.: Privacy-Preserving Channel Access for Internet of Things. IEEE INTERNET THINGS J. 1, 430–445 (2014).
4. Yi, X.: Hash Function Based on Chaotic Tent Maps. IEEE Trans. Circuits Syst. II Express Briefs. 52, 354–357 (2005).
5. Mills, D.L.: Internet Time Synchronization: The Network Time Protocol. IEEE Trans. Commun. 39, 1482–1493 (1991).
6. Yoon, S., Veerarittiphan, C., Sichitiu, M.L.: Tiny-Sync: Tight Time Synchronization for Wireless Sensor Networks. ACM Trans. Sens. Networks. 3, 1–33 (2007).
7. Mallada, E., Meng, X., Hack, M., Zhang, L., Tang, A.: Skewless network clock synchronization. In: 2013 21st IEEE Intern'l Conf. on Network Protocols. pp. 1–10 (2013).
8. Tian, G.S., Tian, Y.C., Fidge, C.: Precise relative clock synchronization for distributed control using TSC registers. J. Netw. Comput. Appl. 44, 63–71 (2014).
9. Shi, L., Jia, C., Lue, S., Liu, Z.: Port and address hopping for active cyber-defense. In: Intelligence and Security Informatics. pp. 295–300 (2007).
10. Shi, L., Jia, C., Lu, S.: DoS evading mechanism upon service hopping. In: 2007 IFIP International Conference on Network and Parallel Comp. Workshop. pp. 119–122 (2007).
11. Zhao, C., Jia, C., Lin, K.: Technique and application of End-hopping in network defense. In: 2010 1st ACIS International Symposium on Cryptography, and Network Security, Data Mining and Knowledge Discovery, E-Commerce and Its Applications, and Embedded Systems. pp. 266–270 (2010).
12. Mittal, P., Kim, D., Hu, Y.-C., Caesar, M.: Mirage: Towards Deployable DDoS Defense for Web Applications. arXiv:1110.1060. (2012).
13. Krylov, V., Kravtson, K.: IP Fast Hopping Protocol Design. In: Proceedings of the 10th Central and Eastern European Software Engineering Conference in Russia (2014).
14. Sifalakis, M., Schmid, S., Hutchison, D.: Network address hopping: a mechanism to enhance data protection for packet communications. In: IEEE International Conference on Communications, 2005. ICC 2005. 2005. pp. 1518–1523 (2005).